\documentclass{ws-procs9x6}

\def\cal{\mathcal}  

\def\ca{C_{\rm A}}

\def\nf{N_{\rm f}\,}

\def\mD{m_{\rm D}}

\def\p{{\bm p}}

\def\q{{\bm q}}
\def\v{{\bm v}}

\def\h{{\bm h}}

\def\F{{\bm F}}

\def\alphas{\alpha_{\rm s}}
\def\gs{g_{\rm s}}

\def\Mrow{Mr\'owczy\'nski}

\def\slashchar#1{\setbox0=\hbox{$#1$}           
   \dimen0=\wd0                                 
   \setbox1=\hbox{/} \dimen1=\wd1               
   \ifdim\dimen0>\dimen1                        
      \rlap{\hbox to \dimen0{\hfil/\hfil}}      
      #1                                        
   \else                                        
      \rlap{\hbox to \dimen1{\hfil$#1$\hfil}}   
      /                                         
   \fi}                                         %

\begin{document}

\title{Transport Coefficients in Hot QCD}

\author{Guy D.~Moore}

\address{Department of Physics, McGill University, \\
	3600 rue University, Montreal, QC H3A 2T8, Canada}

\maketitle

\abstracts{I give a physical explanation of what shear viscosity is,
and what physics determines its value.  Then I explain why determining
the shear viscosity of the Quark-Gluon Plasma is interesting.  I outline
the leading-order calculation of the QGP shear viscosity (and baryon
number diffusion constant), explaining why the quite complicated physics
of parton splitting and Landau-Pomeranchuk-Migdal interference effects
are required for its calculation.  Then I briefly explore the range of
applicability, emphasizing the importance of plasma instabilities.%
}

\section{Introduction}

In this talk I will
discuss work with my collaborators, Peter Arnold and Larry Yaffe, on
transport coefficients in QCD, particularly shear viscosity.

Interest in transport coefficients in the Quark-Gluon Plasma (QGP)
started in the 1980's, and several rough estimates were made at that
time \cite{early_work}.  The first really mature work was by Baym 
{\it et al.\ } \cite{Baymetal} in 1990.  They showed that the dominant
physics involved is small angle scattering, and they demonstrated how to
perform a calculation accurate to leading order in the logarithm of the
strong coupling, $\ln(1/\gs)$.  We started work around 1999 under
the mistaken impression that not much was required to improve their work
to leading order in the coupling $\gs$.  After discovering that even the
leading-log calculation was wrong in the literature, we decided to write two
papers, rather than one--a quick paper \cite{AMY1} computing the leading log
correctly, and a longer one presenting a leading order calculation.
The quick paper took 8 months and 
the ``second'' paper turned out to be six papers
\cite{largeN,AMY2,AMY3,AMY4,AMY5,AMY6}, spanning the next two and a half
years. 

My goal is to explain how the physics of transport coefficients turns
out to be much richer than we had anticipated.  In Section
\ref{sec:whats_shear}, I will explain what shear viscosity is and what
physics is involved; and I will briefly explain why it is an interesting
property of the QGP.  Section \ref{sec:calculation} will outline our
calculation, emphasizing the roles played by soft scatterings, identity
changing scatterings, and collinear processes.  Then, Section
\ref{sec:applicability} will briefly discusses how limited the range of
the calculation is; not only does it depend on weak coupling, but on
a near-equilibrium assumption outside of which the physics is completely
different.

\section{The physics of shear viscosity}
\label{sec:whats_shear}

The velocity of an ideal fluid evolve according to Euler's equations, 
\begin{equation}
\frac{\partial \v}{\partial t} = - \v \!\cdot \! \nabla \, \v
	- \frac{1}{\rho} \nabla p \, , \qquad
p = p(\rho) \, .
\end{equation}
Furthermore, the velocity at a boundary equals
the velocity of that boundary (no slipping), and the stress tensor is
determined by the pressure alone, that is, in the $\v=0$ frame it is
given by 
\begin{equation}
T^{\mu \nu} = (\rho+p) u^\mu u^\nu + p \eta^{\mu \nu} \, ,
\end{equation}
(with $\eta^{\mu \nu}={\rm Diag}[-,+,+,+]$).  In fact, Euler's equations
follow simply from the assumption that the stress tensor will satisfy
this form, from an equation of state, and from stress-energy
conservation, $\partial_\mu T^{\mu \nu} = 0$.

Local equilibrium ensures that these conditions are met.  What enforces
local equilibrium is interactions between degrees of freedom; free
theories never equilibrate.  Therefore, the ideal description is a
better and better approximation as a theory is more and more strongly
coupled; in this case ideality is the opposite of free behavior.

\begin{figure}[ht]
\centerline{
\begin{picture}(300,85)
	\thicklines
	\put(00,5){\line(1,0){240}}
	\put(00,80){\line(1,0){240}}
	\multiput(30,20)(55,0){4}{\vector(1,0){5}}
	\multiput(27,35)(55,0){4}{\vector(1,0){10}}
	\multiput(25,50)(55,0){4}{\vector(1,0){15}}
	\multiput(23,65)(55,0){4}{\vector(1,0){20}}
	\multiput(20,85)(55,0){4}{\vector(1,0){25}}
	\multiput(10,80)(20,0){12}{\line(1,1){10}}
	\multiput(10,-5)(20,0){12}{\line(1,1){10}}
	\put(245,86){Moving top}
	\put(245,76){boundary}
	\put(245,50){Flow varies}
	\put(245,40){with height}
	\put(245,9){Fixed bottom}
	\put(245,0){boundary}
\end{picture}}
\caption{\label{fig1} a system undergoing shear flow}
\end{figure}
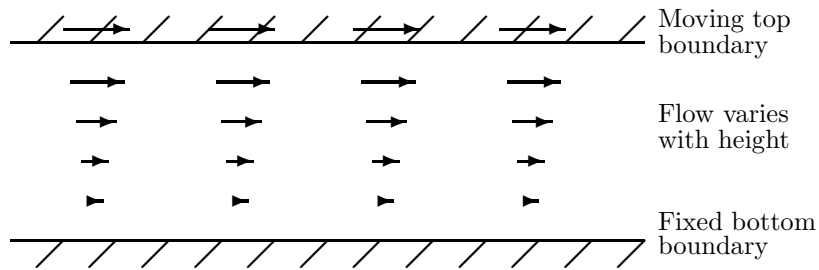

To see this, consider a fluid undergoing shear flow.
Shear flow means that, in the local rest frame, the traceless part of
$\partial_i v_j$ is nonzero.  The simplest
example of shear flow is a fluid trapped between two plates, one fixed
and one moving laterally with velocity $v$, as in Fig.~\ref{fig1};
under ideal Eulerian flow, the fluid at the top is moving with the upper
boundary, with $T_{xz}=0$; therefore there is no lateral force on the
top boundary.

In a real fluid, there will be a ``drag'' force on the top plate; the
shear viscosity is defined by the relation that the force per unit area
is, 
\begin{equation}
\frac{F_x}{A} = T_{xz} = \eta \frac{dv_x}{dz} \, ,
\;\;
\mbox{implying} \;
T_{ij} = p \delta_{ij} + \eta \left( \partial_i v_j {+} \partial_j v_i
	{-} \frac{2}{3} \delta_{ij} \partial_k v_k \right) \, .
\end{equation}

\begin{figure}[ht]
\centerline{
\begin{picture}(300,140)
	\thicklines
	\put(10,10){\line(1,0){280}}
	\put(10,130){\line(1,0){280}}
	\multiput(20,0)(25,0){11}{\line(1,1){10}}
	\multiput(20,130)(25,0){11}{\line(1,1){10}}
	\put(0,67){Initial}
	\put(47,30){\line(1,0){36}}
	\put(65,12){\line(0,1){36}}
	\put(65,30){\oval(30,30)[t]}
	\put(65,30){\oval(26,26)[t]}
	\put(65,30){\oval(28,28)[b]}
	\put(45,70){\line(1,0){40}}
	\put(65,52){\line(0,1){36}}
	\put(68,70){\circle{28}}
	\put(43,110){\line(1,0){44}}
	\put(65,92){\line(0,1){36}}
	\put(71,110){\oval(28,28)[t]}
	\put(71,110){\oval(26,26)[b]}
	\put(71,110){\oval(30,30)[b]}
	\put(85,46){\line(3,2){48}}
	\put(133,78){\vector(1,0){40}}
	\put(85,94){\line(3,-2){48}}
	\put(133,62){\vector(1,0){40}}
	\put(183,70){\line(1,0){44}}
	\put(205,52){\line(0,1){36}}
	\put(205,70){\oval(30,30)[t]}
	\put(205,70){\oval(26,26)[t]}
	\put(211,70){\oval(30,30)[b]}
	\put(211,70){\oval(26,26)[b]}
	\put(240,67){Final}
	%
\end{picture}}
\caption{\label{fig2} Free flow leads to an asymmetrical momentum
distribution.} 
\end{figure}
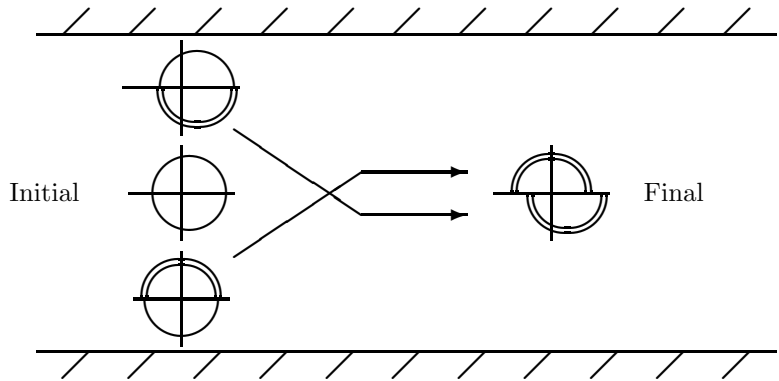

Shear occurs because particles can fly freely.  The
particles near the top of  the fluid region are primarily moving to the
right; those near the bottom are moving right or left with equal
likelihood.  If the particles fly freely, the upward moving particles
near the bottom, and the downward moving particles from the top, will
meet in the middle, as shown in Fig.~\ref{fig2}.
However, the particles coming down from the top tend to be moving to the
right, while particles coming up have no preferred direction; so the
momentum distribution in the center (and everywhere, actually) tends to
become skewed.  Such a skewed distribution has nonvanishing $T_{xz}$ and
there is net transverse momentum flow and a net drag on the top
surface.  Scatterings restore the
distribution towards equilibrium, evening out the momentum
distribution.  The fewer scatterings, the more distorted the momentum
distribution will be.
Therefore, the shear viscosity will be {\em smaller} if
the scattering rate is {\em larger}; $\eta \propto 1/\Gamma$ with
$\Gamma$ the mean scattering rate.
\footnote%
    {%
    Familiar viscous fluids like honey are viscous because the molecules
    get tangled in each other, and are not a good analogy with any
    system which behaves approximately like a collection of freely
    flying objects.  A better example is the air; as the pressure is
    lowered, there are fewer particles available to carry momentum,
    $T_{ij} \propto p$, so
    one would naively expect $\eta \propto p$.  However, at lower
    pressure, collisions are rarer, $\Gamma \propto p$; and so the
    shear viscosity of air is actually independent of air pressure.
    }
The problem of determining the shear viscosity will
be the problem of determining how collision processes re-arrange the
momentum distribution of particles from skewed towards isotropic.

Why is the shear viscosity of the QGP interesting?  One of the main
focuses of this conference is the understanding of nonequilibrium
dynamics; QCD is the most interesting theory to study this and viscosity
and other transport coefficients (baryon number diffusion and electrical
conductivity, which have very similar physics) are theoretically clean
and well defined nonequilibrium dynamical quantities.  That makes them
ideal objects for ``cutting our teeth'' on calculational strategies
for nonequilibrium field theories.

The shear viscosity of QCD is also directly interesting because it gives
a handle on the quality of the hydrodynamical description of heavy ion
collisions; this has been considered at
some length recently by Teaney \cite{Teaney}, who found the corrections
to flow parameters like $v_2$ due to viscosity at RHIC.
Viscosity and particularly diffusion coefficients are also
important in electroweak baryogenesis.

\section{The shear viscosity of the QGP}
\label{sec:calculation}

Shear viscosity describes momentum transfer, and most of the momentum of
a weakly coupled, near-equilibrium QGP is in quasiparticle excitations
with momentum $p \sim 3 T$.  These should be adequately described by
kinetic theory, as I implicitly assumed in the last section's discussion.
In fact, the validity of a kinetic description has essentially been
derived diagrammatically, by Jeon for $\lambda \phi^4$ theory
\cite{Jeon} and by others for gauge theories
\cite{Aarts,Basagoiti}.

The physics of viscosity is the physics of direction changing
scattering, and this is dominated by $2 \leftrightarrow 2$ $t$-channel
gluon exchange \cite{early_work}.  In terms of the spatial momentum $q$
exchanged in the scattering process, such scatterings have a cross
section of form, 
\begin{equation}
d\sigma \propto \frac{d^3 q}{q} \frac{1}{q^4} \, ,
\end{equation}
which is quadratically small $q$ divergent---the familiar Coulomb
divergence.  However, what is important for shear viscosity is how
quickly the particles change direction.
Momentum changes from different scatterings add incoherently; so
\begin{equation}
\int q^2 d\sigma \propto \int q^3 \frac{d^3 q}{q} \frac{1}{q^4}
\sim \int \frac{dq}{q} \, ,
\end{equation}
is the relevant quantity and enough collisions must occur for it to add
up to $|\p|^2$, for a particle to change direction.
This integral is log IR divergent.  There are two key lessons.
\begin{enumerate}
\item
Small $q$ collisions dominate, but only by a logarithm.
\item
Large momentum particles take the longest time to change direction;
since $(\delta \p)^2 \sim |\p|^2$ is required, the time scale for direction
change goes as $\tau \propto |\p|^2$.  Therefore
high energy particles contribute disproportionally to $\eta$.
\end{enumerate}

Baym {\it et al.}\cite{Baymetal} provided the first quantitative
treatment which accounted for both of these points.  To find the shear
flow, they wrote the Boltzmann equation for the nonequilibrium
distribution function $f(x,p)$,
\begin{equation}
\frac{\partial f[p,x,t]}{\partial t} = -\v \cdot \nabla_x f[p,x,t]
	- {\cal C}[f][p,x,t] \, ,
\end{equation}
and linearized it.  That is, they took $f = f_0 + \delta f$ and expanded
the collision integral ${\cal C}$ to first order in $\delta f$.  To this
order, one can write
\begin{equation}
\delta f=f_0(1\pm f_0) \partial_i v_j 
	(\hat p_i \hat p_j -\frac{\delta_{ij}}{3}) \chi(p) \, .
\end{equation}
The collision integral is nonlocal in momentum space, so we have an
integral equation to solve, which in general is nasty.  They showed how
to approach this variationally.  They also showed that the
physics of dynamical screening by the plasma, which is contained within
the Hard Thermal Loops (HTL's) of Braaten and Pisarski \cite{HTL}, is
sufficient to render the Coulomb divergence finite; the would be
logarithmic divergence becomes a large logarithm $\ln(1/g)$.  Further,
at leading order in this logarithm ${\cal C}$ is semilocal in momentum,
which allowed Baym {\it at al.\ } to solve for the
shear viscosity of QCD at leading logarithmic order.

In our first paper
\cite{AMY1}, we showed that, even at leading logarithmic order, one must
also include Compton type processes,

\centerline{
\begin{picture}(200,45)
    \put(0,0){\begin{picture}(50,45)
	\thicklines
	\multiput(2,0)(4,4){4}{\oval(4,4)[tl]}
	\multiput(2,4)(4,4){3}{\oval(4,4)[br]}
	\multiput(16,28)(4,4){4}{\oval(4,4)[tl]}
	\multiput(16,32)(4,4){3}{\oval(4,4)[br]}
	\put(14,14){\circle*{3}}
	\put(14,14){\line(0,1){14}}
	\put(14,28){\circle*{3}}
	\put(14,14){\line(1,-1){14}}
	\put(14,28){\line(-1,1){14}}
    \end{picture}}
    \put(60,7){\begin{picture}(70,30)
	\thicklines
	\multiput(2,0)(4,4){4}{\oval(4,4)[tl]}
	\multiput(2,4)(4,4){3}{\oval(4,4)[br]}
	\multiput(30,14)(4,4){4}{\oval(4,4)[tl]}
	\multiput(30,18)(4,4){3}{\oval(4,4)[br]}
	\put(14,14){\circle*{3}}
	\put(14,14){\line(1,0){14}}
	\put(28,14){\circle*{3}}
	\put(14,14){\line(-1,1){14}}
	\put(28,14){\line(1,-1){14}}
    \end{picture}}
    \put(140,0){\begin{picture}(50,45)
	\thicklines
	\multiput(2,0)(4,4){4}{\oval(4,4)[tl]}
	\multiput(2,4)(4,4){3}{\oval(4,4)[br]}
	\multiput(12,28)(-4,4){4}{\oval(4,4)[tr]}
	\multiput(12,32)(-4,4){3}{\oval(4,4)[bl]}
	\put(14,14){\circle*{3}}
	\put(14,14){\line(0,1){14}}
	\put(14,28){\circle*{3}}
	\put(14,14){\line(1,-1){14}}
	\put(14,28){\line(1,1){14}}
    \end{picture}}
\end{picture}}

\noindent
which are logarithmically IR divergent.  The role of these processes is
not to change particle direction, but to change particle {\em identity};
a quark on scattering becomes a gluon and {\it vice versa}.  Because of
their different group Casimir, gluons equilibrate more quickly than
quarks; so the conversion of hard to equilibrate quarks into easy to
equilibrate gluons accelerates thermalization.  This is needed at
leading-log order, though it is only numerically important for
electrical conductivity in QED.  With these processes, a leading-log
treatment is possible \cite{AMY1}, yielding, for instance,
\begin{eqnarray}
\sigma & = & \frac{12^4 \zeta^2(3) \pi^{-3} N_{lept}}
	{3\pi^2 + 32 N_{spec}} \; \frac{T}{e^2 \ln 1/e} \, , \\
D_{\rm quark} & = & \frac{2^4 3^6 \zeta^2(3) \pi^{-3}}
	{24 + 4N_{\rm f} + \pi^2} \; \frac{1}{g^4 T \ln 1/g} \, .
\end{eqnarray}
The result for $\eta$ is more cumbersome and can be found in our paper
\cite{AMY1}.  The above results are only accurate to $0.5\%$ because
they were made with the {\it Ansatz} $\xi(p) \propto p^2$.
More accurate results are presented in the paper.

What is required to convert these to a leading-order computation?
One only needs to treat the collision integral more carefully,
both in the $q \sim T$ momentum region where the collision integral is
nonlocal, and in the $q \sim \mD \sim \gs T$ region where screening
effects must be dealt with.  For heavy quark diffusion, this is
sufficient; Derek Teaney and I recently derived a complete leading order
result;
\begin{equation}
D = \frac{72\pi}{C_{\rm f} \gs^4 T}
\left[ \left(N_{\rm c} {+} \frac{\nf}{2} \right) \left( \ln \frac{2 T}{m_D}
       {+} \frac{1}{2} - \gamma_E
       {+} \frac{\zeta'(2)}{\zeta(2)} \right) + \frac{\nf \ln 2}{2}
\right]^{-1} \! .
\end{equation}

However, viscosity in the QGP and light quark diffusion turn out to
involve more physics, and therefore more work; one must also include
inelastic processes with collinear external states.

\centerline{
\begin{picture}(150,45)
    \put(0,0){\begin{picture}(60,40)
	\thicklines
	\put(0,0){\line(2,1){20}}
	\put(20,10){\circle*{3}}
	\put(20,10){\line(2,-1){20}}
	\multiput(20,12)(0,8){3}{\oval(4,4)[r]}
	\multiput(20,16)(0,8){2}{\oval(4,4)[l]}
	\put(20,30){\circle*{3}}
	\put(0,40){\line(2,-1){20}}
	\put(20,30){\line(2,1){20}}
    \end{picture}}
    \put(53,5){\begin{picture}(30,30)
	\thicklines
	\multiput(0,13)(0,4){2}{\line(1,0){15}}
	\put(14,20){\line(2,-1){10}}
	\put(14,10){\line(2,1){10}}
    \end{picture}}
    \put(90,0){\begin{picture}(60,40)
	\thicklines
	\put(0,0){\line(2,1){20}}
	\put(20,10){\circle*{3}}
	\put(20,10){\line(2,-1){20}}
	\multiput(20,12)(0,8){3}{\oval(4,4)[r]}
	\multiput(20,16)(0,8){2}{\oval(4,4)[l]}
	\put(20,30){\circle*{3}}
	\put(0,40){\line(2,-1){20}}
	\put(20,30){\line(2,1){20}}
	\put(30,5){\circle*{3}}
	\multiput(30,7)(4,4){3}{\oval(4,4)[br]}
	\multiput(34,7)(4,4){2}{\oval(4,4)[tl]}
    \end{picture}}
\end{picture}}
\noindent
Particles in the plasma undergo the left process at an $O(\alphas T)$
rate, not $\alphas^2 T$, because of the Coulomb divergence already
discussed.  The right process is slower by a factor of $\alphas$; this
is true even when the radiated gluon has momentum $O(T)$, provided it is
collinear with the particle which emits it.  The rate for this process
is therefore as large as the large angle scattering rate.

We saw previously that the particles furthest from equilibrium were
those with the most energy, $\delta f \propto f |\p|^2$.  Collinear
processes like the one just presented tend to split these high energy
particles into lower energy ones, which are more easily bent; therefore
they lead to equilibration and lower $\eta$, even though they do not
change particle direction.  Numerically, we find that they
reduce $\eta$ by about $10\%$ \cite{AMY6}.

But life is even more complicated; one cannot ignore interference
between emission amplitudes of form

\centerline{\begin{picture}(260,45)
    \put(0,0){\begin{picture}(100,45)
	\thicklines
	\put(0,25){\line(1,0){90}}
	\multiput(20,25)(25,0){3}{\circle*{3}}
	\multiput(10,0)(25,0){3}{\line(2,1){10}}
	\multiput(20,5)(25,0){3}{\line(2,-1){10}}
	\multiput(20,5)(25,0){3}{\circle*{3}}
	\multiput(20,7)(0,8){3}{\oval(4,4)[r]}
	\multiput(20,11)(0,8){2}{\oval(4,4)[l]}
	\multiput(45,7)(0,8){3}{\oval(4,4)[r]}
	\multiput(45,11)(0,8){2}{\oval(4,4)[l]}
	\multiput(70,7)(0,8){3}{\oval(4,4)[r]}
	\multiput(70,11)(0,8){2}{\oval(4,4)[l]}
	\multiput(8,25)(4,4){3}{\oval(4,4)[tl]}
	\multiput(8,29)(4,4){3}{\oval(4,4)[br]}
    \end{picture}}
    \put(101,10){\Large $+\ldots +$}
    \put(170,0){\begin{picture}(100,45)
	\thicklines
	\put(0,25){\line(1,0){90}}
	\multiput(20,25)(25,0){3}{\circle*{3}}
	\multiput(10,0)(25,0){3}{\line(2,1){10}}
	\multiput(20,5)(25,0){3}{\line(2,-1){10}}
	\multiput(20,5)(25,0){3}{\circle*{3}}
	\multiput(20,7)(0,8){3}{\oval(4,4)[r]}
	\multiput(20,11)(0,8){2}{\oval(4,4)[l]}
	\multiput(45,7)(0,8){3}{\oval(4,4)[r]}
	\multiput(45,11)(0,8){2}{\oval(4,4)[l]}
	\multiput(70,7)(0,8){3}{\oval(4,4)[r]}
	\multiput(70,11)(0,8){2}{\oval(4,4)[l]}
	\multiput(80,25)(4,4){3}{\oval(4,4)[tl]}
	\multiput(80,29)(4,4){3}{\oval(4,4)[br]}
    \end{picture}}
\end{picture}}
\noindent
The effect of such interference is referred to as the
Landau-Pomeranchuk-Migdal (LPM) effect \cite{LPM}, and it suppresses the
efficiency of collinear emission.  To see the physical origin of the
suppression, consider a particle undergoing multiple scatterings, and
remember that it possesses a finite transverse wave packet.  That wave
packet continues to physically overlap the wave packet of the emitted
radiation for some time, called the formation time; if there is a second
scattering in that time, the two radiation fields overlap and interfere; 

\centerline{
\epsfxsize=0.7\textwidth
\epsfbox{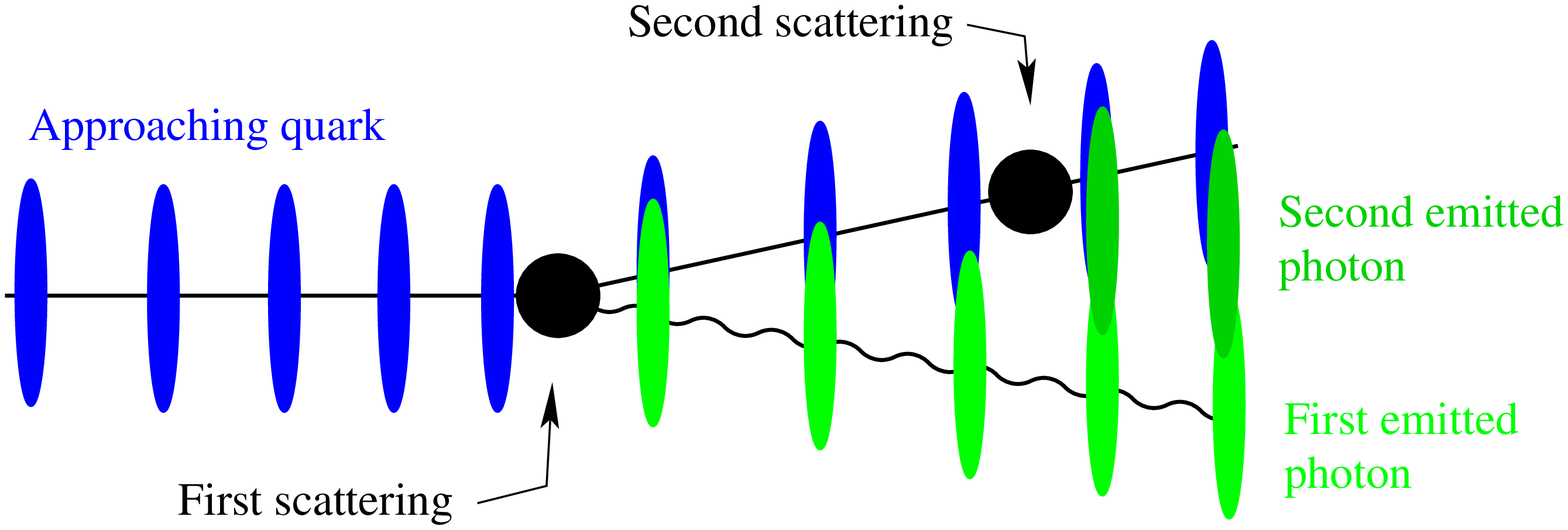}}
\noindent
The radiations sum in amplitude; on average the interference
is destructive.

Dealing with this complication takes a lot of work, and has been the
topic of several studies \cite{others,AMY2,AMY4}.
The upshot is that the rate at which a particle of momentum $p$
undergoes a soft scattering which induces a bremsstrahlung of a gluon of
energy $k$, is
\begin{eqnarray}
\label{eq:dGamma}
\frac{d\Gamma(p,k)}{dk dt} & = & \frac{C_s \gs^2}{16\pi p^7} 
        \frac{1}{1 \pm e^{-k/T}} \frac{1}{1 \pm e^{-(p-k)/T}} \times
\left\{ \begin{array}{cc} 
        \frac{1+(1{-}x)^2}{x^3(1{-}x)^2} & q \rightarrow qg \\
        N_{\rm f} \frac{x^2+(1{-}x)^2}{x^2(1{-}x)^2} & g \rightarrow qq \\
        \frac{1+x^4+(1{-}x)^4}{x^3(1{-}x)^3} & g \rightarrow gg \\
        \end{array} \right\} 
	\times \nonumber \\ && \times
\int \frac{d^2 \h}{(2\pi)^2} 2 \h \cdot {\rm Re}\: \F(\h,p,k) \, ,
\end{eqnarray}
where $\F$ is the solution of the following integral equation:
\begin{eqnarray}
&& 2\h = i \delta E \F(\h) + g^2 
	\!\!\int \!\! \frac{d^2 \q_\perp}{(2\pi)^2}
	C(\q_\perp)\Big\{ 
	\frac{2C_s {-}\ca}{2}
	[\F(\h)-\F(\h{-}k\,\q_\perp)] +
        \nonumber \\ && \hspace{0.8in}
        + \frac{\ca}{2}[\F(\h)-\F(\h{+}p\,\q_\perp) 
        + \F(\h)-\F(\h{-}(p{-}k)\,\q_\perp)] \Big\} , 
\label{eq:integral_eq1}
        \nonumber \\
&&\delta E \equiv
        \frac{\h^2}{2pk(p{-}k)} { + \frac{m_k^2}{2k} +
        \frac{m_{p{-}k}^2}{2(p{-}k)} - \frac{m_p^2}{2p}} \, ,
\label{eq:integral_eq2}
\end{eqnarray}
with 
$C(\q_\perp) = \frac{\mD^2}{ \q_\perp^2(\q_\perp^2{+}\mD^2)}$,
$\mD^2 = \frac{\gs^2 T^2}{6} (2 N_{\rm c} {+} N_{\rm f})$.
Determining $d\Gamma/dkdt$ therefore requires the solution of an
integral equation; and this for a collision term which, to solve the
Boltzmann equation, must appear as a coefficient in an integral equation
which is to be solved.  However, the numerics turn out not to be very
demanding.

\begin{figure}
\centerline{
\epsfxsize=0.45\textwidth \epsfbox{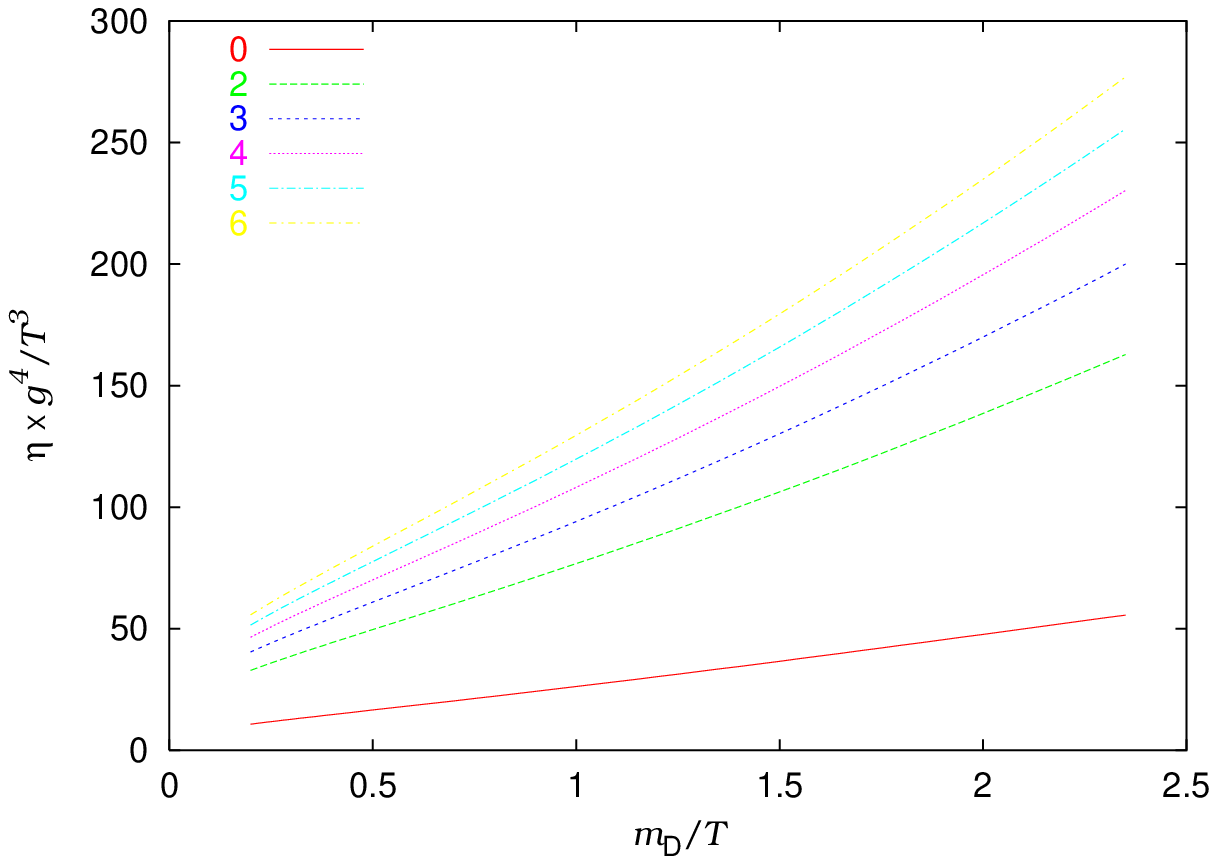}
\hspace{0.05\textwidth}
\epsfxsize=0.45\textwidth \epsfbox{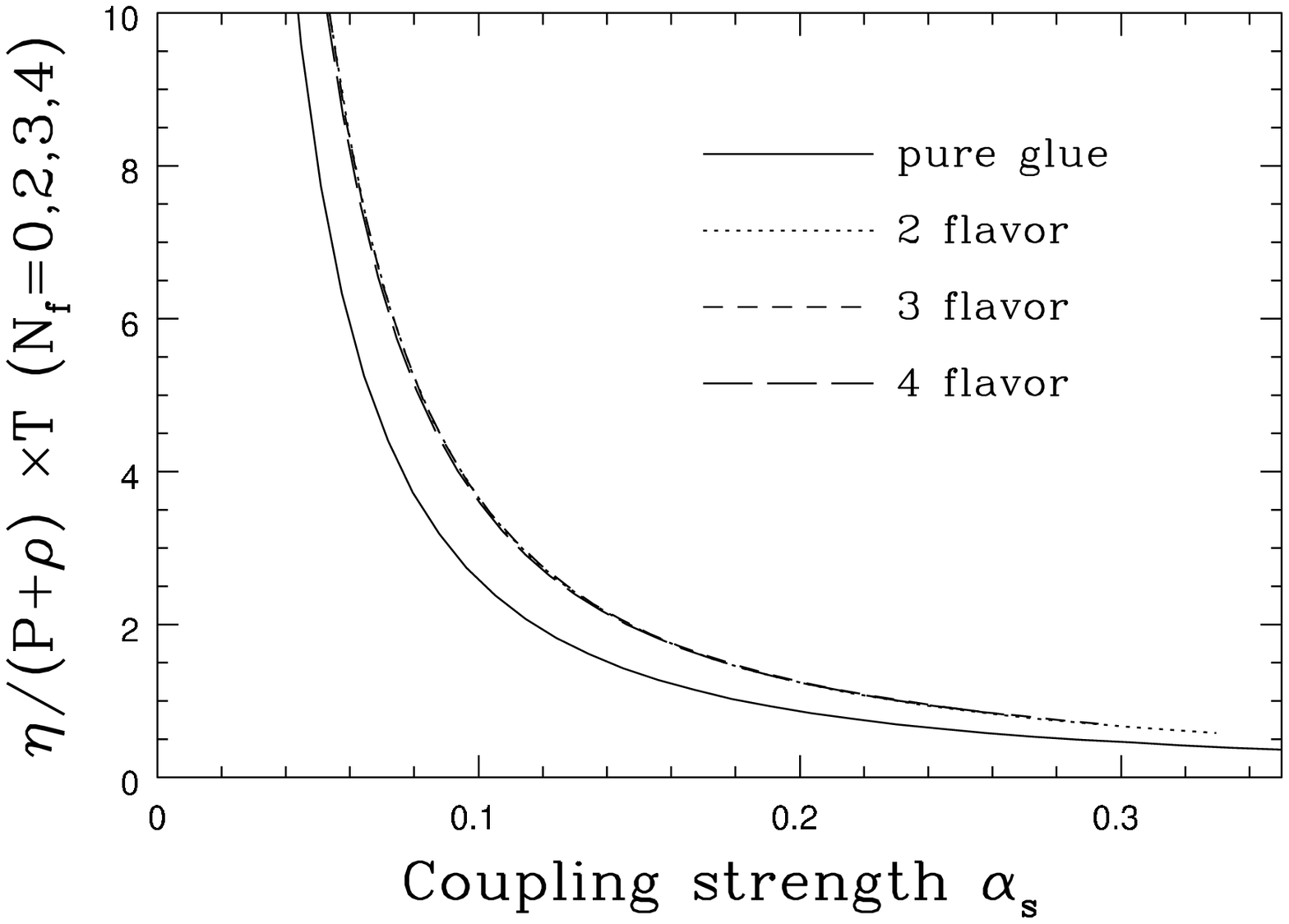}
}
\caption{\label{fig3}Shear viscosity, scaled by $\alphas^{-2}$ (left)
and by enthalpy density (right).}
\end{figure}

These ingredients have been combined into a kinetic theory for the QGP
\cite{AMY5}, which has been linearized and solved to determine the
transport coefficients at leading order in the strong coupling
\cite{AMY6}.  Results for shear viscosity are presented in
Fig.~\ref{fig3}.  At
$\alphas=1/3$, $\eta$ is quite small, and hydrodynamics should be
nearly ideal; but the perturbative computation is breaking down at this
point and the results cannot be trusted (factor of 3 uncertainty?).

\section{Range of applicability of QGP transport coefficients}
\label{sec:applicability}

Two key assumptions underlie the above calculation.  The first is that the
coupling is actually weak, at the scale relevant to the physics in
question.  For electroweak phase transition applications this is
probably a good approximation; for RHIC it is dubious, though most of
the theory community is guilty of this assumption.  I will not discuss
this more here.

The other assumption underpinning our treatment is, that the system is
near equilibrium.  One can check self-consistency by comparing the time
scale of relaxation with the age of the plasma.  The ratio
$\eta / (\rho + p)$ is a time or length scale, called the sound attenuation
length; it differs by a geometric factor $\sim 1/5$ from the relaxation
time of the plasma.  At $\alphas=1/3$, Fig.~\ref{fig3} shows that this
is $\sim 2/3T$, so $\tau \sim 3/T$, which is a few fermis at RHIC.
Therefore we expect hydrodynamic treatments to be valid after $\sim 3$
fermi at RHIC.  One might question a hydro description for times earlier
than that.

How different could the early time physics be?  The answer turns out to
be, pretty different.  For systems which are far from equilibrium but
approximately isotropic, kinetic theory should be applicable under fairly
weak conditions \cite{AMY5}.  But for systems with strong momentum space
anisotropy, this turns out not to be the case.  A treatment involving
soft gauge fields becomes essential, because these soft gauge fields
show exponential growth due to plasma instabilities \cite{Weibel}.  It
has recently been noted that this may be important for QGP physics
\cite{Mrow,Strickland1,ALM}, and it is the topic of two other talks at
this conference \cite{Strickland2,Peter}.  The short story is that, for
plasmas which are far from isotropic, soft gauge field modes can grow at
a rate suppressed by only a single power of $\gs$, and may \cite{AL}
randomize the directions of other particles on this time scale.

\section{Conclusions}
\label{sec:conclusions}

Transport coefficients in gauge theories can be computed within
kinetic theory, but they involve rich physics; hard
thermal loops, multiple scattering processes, collinear splitting.  The
kinetic description is only possible after resumming certain multiple
scattering emission processes to account for LPM interference.  It has
taken nearly 20 years to go from the first attempts at a quantitative
analysis to one which is complete at leading order in the coupling
$\gs$. 

The calculation of transport coefficients has been directly useful to
the heavy ion community \cite{Teaney}, but it has also been a good
object lesson in non-equilibrium field theory in QCD.  Most importantly,
we have learned just how narrow the domain of validity of kinetic theory
is, and how much more complicated the physics of highly anisotropic
plasmas may be.  The problem of understanding  plasma
instabilities in QCD is only beginning to be addressed and should be the
subject of much future work.

\section*{Acknowledgements}

I particularly thank my co-workers, Peter Arnold and Larry Yaffe.  I
also thank Mike Strickland, Stan \Mrow, and Francois Gelis for
conversations which have been useful in this work over the past years.
My work is currently supported by grants from the
Natural Sciences and Engineering Research Council of Canada and by le Fonds
Qu\'eb\'ecois de la Recherche sur la Nature et les Technologies.


\begin{thebibliography}{0}

\bibitem{early_work}
A.~Hosoya, M.~Sakagami and M.~Takao,
Annals Phys.\  {\bf 154}, 229 (1984);
A.~Hosoya and K.~Kajantie,
Nucl.\ Phys.\  {\bf B250}, 666 (1985).

\bibitem{Baymetal}
G.~Baym, H.~Monien, C.~J.~Pethick and D.~G.~Ravenhall,
Phys.\ Rev.\ Lett.\  {\bf 64}, 1867 (1990);
Nucl.\ Phys.\  {\bf A525}, 415C (1991).

\bibitem{AMY1}
P.~Arnold, G.~D.~Moore and L.~G.~Yaffe,
JHEP {\bf 0011}, 001 (2000)
[hep-ph/0010177].

\bibitem{largeN}
G.~D.~Moore,
JHEP {\bf 0105}, 039 (2001)
[hep-ph/0104121].

\bibitem{AMY2}
P.~Arnold, G.~D.~Moore and L.~G.~Yaffe,
JHEP {\bf 0111}, 057 (2001)
[hep-ph/0109064].

\bibitem{AMY3}
P.~Arnold, G.~D.~Moore and L.~G.~Yaffe,
JHEP {\bf 0112}, 009 (2001)
[hep-ph/0111107].

\bibitem{AMY4}
P.~Arnold, G.~D.~Moore and L.~G.~Yaffe,
JHEP {\bf 0206}, 030 (2002)
[hep-ph/0204343].

\bibitem{AMY5}
P.~Arnold, G.~D.~Moore and L.~G.~Yaffe,
JHEP {\bf 0301}, 030 (2003)
[hep-ph/0209353].

\bibitem{AMY6}
P.~Arnold, G.~D.~Moore and L.~G.~Yaffe,
JHEP {\bf 0305}, 051 (2003)
[hep-ph/0302165].

\bibitem{Teaney}
D.~Teaney,
Phys.\ Rev.\ C {\bf 68}, 034913 (2003)
[nucl-th/0301099].

\bibitem{Jeon}
S.~Jeon,
Phys.\ Rev.\ D {\bf 52}, 3591 (1995)
[hep-ph/9409250];
S.~Jeon and L.~G.~Yaffe,
Phys.\ Rev.\ D {\bf 53}, 5799 (1996)
[hep-ph/9512263].

\bibitem{Aarts}
G.~Aarts and J.~M.~Martinez Resco,
JHEP {\bf 0211}, 022 (2002)
[hep-ph/0209048];
G.~Aarts and J.~M.~Martinez Resco,
Phys.\ Rev.\ D {\bf 68}, 085009 (2003)
[hep-ph/0303216].

\bibitem{Basagoiti}
M.~A.~Valle Basagoiti,
Phys.\ Rev.\ D {\bf 66}, 045005 (2002)
[hep-ph/0204334].

\bibitem{HTL}
E.~Braaten and R.~D.~Pisarski,
Nucl.\ Phys.\ B {\bf 337}, 569 (1990).

\bibitem{MT}
D.\ Teaney and G.\ D.\ Moore, in preparation.

\bibitem{LPM}
L.~D.~Landau and I.~Pomeranchuk,
Dokl.\ Akad.\ Nauk Ser.\ Fiz.\  {\bf 92} (1953) 535;
Dokl.\ Akad.\ Nauk Ser.\ Fiz.\  {\bf 92} (1953) 735;
A.~B.~Migdal, Dokl.\ Akad.\ Nauk S.S.S.R.~{\bf 105}, 77 (1955);
Phys.\ Rev.\  {\bf 103}, 1811 (1956).

\bibitem{others}
See for instance,
R.~Blankenbecler and S.~D.~Drell,
Phys.\ Rev.\ D {\bf 53}, 6265 (1996);
R.~Baier, Y.~L.~Dokshitzer, A.~H.~Mueller, S.~Peigne and D.~Schiff,
Nucl.\ Phys.\ B {\bf 478}, 577 (1996)
[hep-ph/9604327];
B.~G.~Zakharov,
Phys.\ Atom.\ Nucl.\  {\bf 61}, 838 (1998)
[Yad.\ Fiz.\  {\bf 61}, 924 (1998)]
[hep-ph/9807540].

\bibitem{Weibel}
E. S. Weibel,
Phys.\ Rev.\ Lett.\ {\bf 2}, 83 (1959).

\bibitem{Mrow}
S.~\Mrow,
Phys.\ Lett.\ B {\bf 214}, 587 (1988);
Phys.\ Lett.\ B {\bf 314}, 118 (1993);
Phys.\ Lett.\ B {\bf 393}, 26 (1997)
[hep-ph/9606442].

\bibitem{Strickland1}
P.~Romatschke and M.~Strickland,
Phys.\ Rev.\ D {\bf 68}, 036004 (2003)
[hep-ph/0304092];
hep-ph/0406188.

\bibitem{ALM}
P.~Arnold, J.~Lenaghan and G.~D.~Moore,
JHEP {\bf 0308}, 002 (2003)
[hep-ph/0307325].

\bibitem{Strickland2}
P.~Romatschke and M.~Strickland,
these proceedings,
hep-ph/0408314.

\bibitem{Peter}
Peter Arnold, these proceedings.

\bibitem{AL}
P.~Arnold and J.~Lenaghan,
hep-ph/0408052.

\end{thebibliography}
\end{document}